\documentclass[journal]{IEEEtran}
\usepackage{verbatim}
\ifCLASSINFOpdf
\else
   \usepackage[dvips]{graphicx}
\fi
\usepackage{multirow}
\usepackage{booktabs}
\usepackage{caption}
\usepackage{graphicx}
\usepackage{epstopdf}
\usepackage{subfigure}
\usepackage{amsmath,setspace}
\usepackage{amssymb}
\usepackage{amsthm}
\usepackage{amsfonts}
\usepackage[linesnumbered,boxed,ruled,commentsnumbered]{algorithm2e}
\usepackage{algpseudocode}
\usepackage{array}
\usepackage{cite}
\usepackage{url}
\usepackage{hyperref}
\usepackage{color}
\usepackage{epstopdf}
\usepackage{overpic}
\usepackage{lipsum}
\usepackage{cuted}
\usepackage{stfloats}
\definecolor{myred}{RGB}{234, 131, 121}

\theoremstyle{noparens}

\begin{document}

\title{Deep Learning Based Stage-wise \\Two-dimensional Speaker Localization \\with Large Ad-hoc Microphone Arrays}

\author{Shupei Liu$^\dagger$, Linfeng Feng$^\dagger$, Yijun Gong, Chengdong Liang, Chen Zhang, \\
  Xiao-Lei Zhang, \textit{Senior Member, IEEE}, and Xuelong Li, \textit{Fellow, IEEE}
\thanks{Shupei Liu and Linfeng Feng contributed equally to this work.}
\thanks{Xiao-Lei Zhang is the corresponding author.}
\thanks{Shupei Liu, Linfeng Feng, Yijun Gong, Chengdong Liang, Chen Zhang and Xiao-Lei Zhang are with the School of Marine Science and Technology, Northwestern Polytechnical University, Xi'an 710072, China (e-mail: shupei.liu@mail.nwpu.edu.cn; fenglinfeng@mail.nwpu.edu.cn; gongyj@mail.nwpu.edu.cn; liangchengdong@mail.nwpu.edu.cn; chen7zhang@mail.nwpu.edu.cn; xiaolei.zhang@nwpu.edu.cn).}
\thanks{Xuelong Li is with the Institute of Artificial Intelligence (TeleAI), China Telecom Corp Ltd, 31 Jinrong Street, Beijing 100033, P. R. China (e-mail: li@nwpu.edu.cn).}
}

\markboth{}
{Shell \MakeLowercase{\textit{et al.}}: Bare Demo of IEEEtran.cls for IEEE Journals}
\maketitle

\begin{abstract}
While deep-learning-based speaker localization has shown advantages in challenging acoustic environments, it often yields only direction-of-arrival (DOA) cues rather than precise two-dimensional (2D) coordinates. To address this, we propose a novel deep-learning-based 2D speaker localization method leveraging ad-hoc microphone arrays, where an ad-hoc microphone array is composed of randomly distributed microphone nodes, each of which is equipped with a traditional array. Specifically, we first employ convolutional neural networks at each node to estimate speaker directions. Then, we integrate these DOA estimates using triangulation and clustering techniques to get 2D speaker locations. To further boost the estimation accuracy, we introduce a node selection algorithm that strategically filters the most reliable nodes. Extensive experiments on both simulated and real-world data demonstrate that our approach significantly outperforms conventional methods. The proposed node selection further refines performance. The real-world dataset in the experiment, named Libri-adhoc-node10 which is a newly recorded data described for the first time in this paper, is online available at \href{https://github.com/Liu-sp/Libri-adhoc-nodes10}{https://github.com/Liu-sp/Libri-adhoc-nodes10}.
\end{abstract}



\IEEEpeerreviewmaketitle

\section{Introduction}
\IEEEPARstart{S}peaker localization aims to localize speaker positions using speech signals recorded by microphones. It finds wide applications in sound event detection and localization \cite{bai20233d},  speaker separation \cite{jiang2014binaural,taherian2022multi,fu2023locate} and diarization \cite{wang2022localization,taherian2023multi,gburrek2023spatial}, etc.

\subsection{Motivation and challenges}
Speaker localization in adverse acoustic environments with strong reverberation and noise interference is challenging. Conventional speaker localization requires obtaining the directions of speech sources, also known as direction-of-arrival (DOA) estimation. Representative methods include multiple signal classification (MUSIC) \cite{schmidt1986multiple} and steered response power with phase transform (SRP-PHAT) \cite{dibiase2000high}.

Recently, with the rapid development of deep-learning-based speech separation and enhancement \cite{wang2018supervised}, deep-learning-based DOA estimation has received increasing attention \cite{xiao2015learning,wang2018robust,chakrabarty2019multi,nguyen2020robust,qian2021multi,he2021neural,qian2022deep,mack2022signal,songgong2022acoustic}. Some methods utilize deep models to estimate noise-robust variables that are then fed into conventional DOA estimators \cite{wang2018robust}. Other methods formulate DOA estimation as a classification problem of azimuth classes \cite{xiao2015learning}. Spatial acoustic features like generalized cross correlation \cite{xiao2015learning}, phase spectrograms \cite{chakrabarty2019multi}, spatial pseudo-spectrum \cite{nguyen2020robust}, and circular harmonic features \cite{songgong2022acoustic} are frequently extracted as input to deep models. Convolutional neural networks (CNNs) are popular in the study of the DOA estimation \cite{chakrabarty2019multi,nguyen2020robust,he2021neural,mack2022signal}. Following the above directions, many generalized issues were explored \cite{nguyen2020robust,he2021neural,qian2021multi,qian2022deep}, exhibiting improved performance over conventional approaches.

However, in many applications, obtaining a speaker's 2-dimensional (2D) or 3-dimensional coordinate is more helpful than merely obtaining the DOA. Ad-hoc microphone arrays may be able to address the problem. An ad-hoc microphone array is a group of randomly distributed cooperative microphone nodes, each of which contains a traditional microphone array, like a uniform linear array. The advantages of ad-hoc microphone arrays lie in that (i) they can be easily deployed and widespread in real world by organizing online devices, and (ii) they can reduce the occurrence probability of far-field speech signal processing \cite{zhang2021deep}. As for the sound-source localization problem, analogous to prior investigations such as \cite{chen2021scaling,yang2022deep,liang2022multi}, whether an ad-hoc array can substantially outperform traditional fixed arrays by handling a substantial multitude of nodes, needs a deep investigation.

\begin{figure*}[t]
\centering
\includegraphics[width=0.92\textwidth]{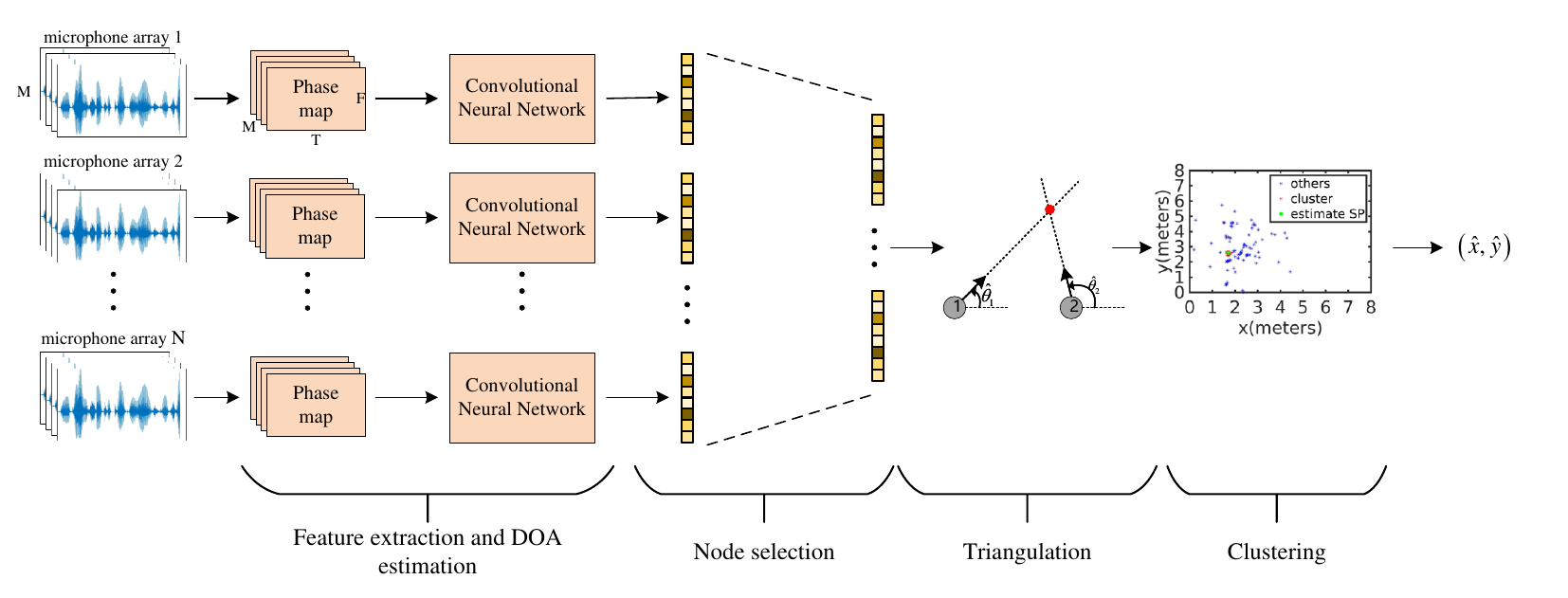}
\caption{Diagram of the proposed 2-dimensional speaker localization method based on deep learning.}
\label{fig1}
\end{figure*}

Conventional sound source localization approaches based on ad-hoc microphone arrays primarily employ signal processing methods, as described in \cite{cobos2017survey}. Recent progress in speaker localization leverages deep learning in conjunction with distributed microphone nodes \cite{sun2017indoor,vesperini2018localizing,le2019learning,kindt20212d,hahmann2022sound,gong2022end,yan2023indoor}, which is the focus of this paper. For example, \cite{vesperini2018localizing} utilize multiple deep-learning-based nodes to directly predict 2D speaker coordinates. Alternatively, \cite{sun2017indoor,le2019learning,gong2022end,yan2023indoor} formulate indoor localization as a spatial grid classification problem. \cite{kindt20212d} derives 2D coordinates through triangulation of two distributed nodes. \cite{hahmann2022sound} feeds DOA estimates from each node into a deep neural network (DNN) to obtain the final speaker location.

While these pioneering works highlight the potential of deep learning techniques, their investigations have been limited to sparse node numbers (e.g. two nodes) and additional constraints. These constraints include fixed node positioning for both training and testing in the same room \cite{sun2017indoor,vesperini2018localizing,le2019learning,hahmann2022sound,yan2023indoor}. Alternatively, \cite{kindt20212d} mandates identical spatial node patterns for training and testing, making it difficult to maximize the flexibility of ad-hoc arrays. Additionally, \cite{gong2022end} is tailored for scenarios where each node consists of a single microphone, precluding integration with prevalent DOA estimation techniques.

\subsection{Framework of the proposed method}\label{subsec:framework}
In pursuit of the flexibility and advantages of ad-hoc arrays, this paper introduces a deep-learning-based 2D speaker localization leveraging large-scale ad-hoc microphone arrays. The framework of the proposed method is shown in Fig.~\ref{fig1}. Specifically, it comprises a feature extraction module, a DOA estimation module, a node selection algorithm, and a triangulation and clustering method. The DOA estimation module provides speaker directions. The node selection algorithm selects ad-hoc nodes that yield highly reliable DOA estimates. The triangulation module yield a rough 2D speaker location from any two randomly selected ad-hoc nodes. At last, the clustering algorithm conducts clustering on all rough speaker locations, and takes the clustering center as the final accurate speaker location.

\subsection{Goals and contributions}

The novelty and contributions of the proposed method lie in that:
\begin{itemize}
  \item \textbf{We have proposed a stage-wise deep-learning-based 2D sound source localization method.} The method has been described in Section \ref{subsec:framework}. It does not require the ad-hoc nodes to be at fixed positions. It is a stage-wise framework, which is flexible in incorporating many advanced techniques in DOA estimations, node selection strategies, and clusterings. At last, it bridges the gap between conventional signal processing methods and recent deep learning methods.

  \item
  \textbf{We have employed an advanced classification-based DOA estimation algorithm that is free of quantization errors.} The backbone network is CNN, where a mask layer is used to enhance the robustness of the DOA estimation. Furthermore, to improve the accuracy of the DOA estimation of the CNN-based classification model, we incorporate a quantization-error-free soft label encoding and decoding strategy.

  \item
  \textbf{We have recorded a real-world dataset named Libri-adhoc-nodes10.}  The Libri-adhoc-nodes10 dataset is a 432-hour collection of replayed speech of the ``test-clean'' subset of the Librispeech corpus \cite{panayotov2015librispeech}, where an ad-hoc microphone array with 10 nodes were placed in an office and a conference room respectively. Each node is a linear array of four microphones. For each room, 4 array configurations with 10 distinct speaker positions per configuration were designed.
\end{itemize}

Experimental results on both simulated data and real-world data demonstrate the superiority of the proposed method over existing approaches. Moreover, the models trained on simulated data perform well on real-world test data.

This paper is organized as follows. Section~\ref{sec:doa} describes the DOA estimation algorithm based on CNN at each single ad-hoc node. Section~\ref{sec:inter} describes the process on how to integrates the DOA estimations of all ad-hoc nodes into a 2D position estimate. Section~\ref{sec:data} describes the collected Libri-adhoc-node10 dataset. Section~\ref{sec:exp} demonstrates the advantages of the proposed method on both simulated and real-world data. Section~\ref{sec:dis} discusses some limitations of the proposed method both theoretically and empirically. Finally, Section~\ref{sec:conclusion} concludes our findings.

\section{CNN-based DOA estimation at each single ad-hoc node} \label{sec:doa}

In this section, we first describe the CNN backbone networks in Section \ref{subsec:backbone}. Then, we discuss the permutation ambiguity problem of multi-source localization training in Section~\ref{subsec:permutation}. Section~\ref{subsec:uld} introduces our solution, named unbiased label distribution encoding, to the quantization error problem. Finally, in Section~\ref{subsec:wad}, we describe the soft decoding, which transforms the DNN output to a DOA estimate.

\subsection{Backbone networks}\label{subsec:backbone}

This subsection describes two backbone networks for the multi-speaker DOA estimation problem. The first one is a modified classic CNN-based multi-label classification (CNN-MLC) network \cite{chakrabarty2019multi}. The second one is a recent CNN-based masking (CNN-Mask) network \cite{subramanian2022deep}.

\subsubsection{CNN-MLC}
Consider a room with an ad-hoc microphone array of $N$ nodes and $B$  speakers, where each node comprises a conventional array of $M$ microphones.

The short-time Fourier transform (STFT) of a speech recording at the $i$-th microphone of an ad-hoc node is $Y_{i}(t, f)=A_{i}(t, f) e^{j \phi_{i}(t, f)}$,
where $Y_i(t,f)$ is the STFT at the $t$-th frame and $f$-th frequency bin, and $A_{i}(t, f)$ and $\phi_{i}(t, f)$ are the magnitude and phase components of the STFT respectively, $\forall i \in {1,\dots,M}$, $\forall t \in {1,\dots,T}$, and $\forall f \in {1,\dots,F}$ where $T$ is the number of the time frames of the recording and $F$ is the number of frequency bins. We group the phase spectrograms of all microphones of the node into a $M \times F \times T$ matrix, denoted as $\mathbf{\Phi}$, i.e. $\mathbf{\Phi} = [\phi_{i}(t, f)]_{i,f,t}\in \mathbb{R}^{M \times F \times T}$.

The deep-learning-based DOA estimation is formulated as a classification problem of $L+1$ azimuth angles in the full azimuth range, where the multi-speaker DOA problem is formulated as a multi-label classification (MLC) problem \cite{chakrabarty2019multi}. Table \ref{tab:cnn-mlc} describes the architecture of the CNN-MLC, which is a modified version of \cite{chakrabarty2019multi}.

After taking $\mathbf{\Phi}$ into the CNN-MLC, the predicted distribution of the CNN-MLC can be represented as $\hat{\boldsymbol{\rho}} \in [0, 1]^{L+1}$, where $\hat \rho_{l}$ denotes the probability of the speaker being in the $l$-th azimuth class of the DOA, {$\forall l \in {0,\dots,L}$}. This can be formulated as:
\begin{equation}\label{eq:xx}
\hat{\boldsymbol{\rho}}= \mbox{CNN}(\mathbf{\Phi})
\end{equation}
where the $B$ classes with the highest probabilities are the estimated DOAs of the $B$ speakers.

\renewcommand\arraystretch{1.3}
\begin{table}[t!]
  \centering
  \caption{Architecture of the CNN-MLC \cite{chakrabarty2019multi}.}
  \label{tab:cnn-mlc}
  \scalebox{0.92}{\begin{tabular}{c c c}
  \hline
  Layer name &Structure & Output size \\
  \hline
   Input           & ---    &   $1 \times 4 \times 256$    \\

   Conv-1        & $2 \times 1$, Stride=(1, 1)  & $4 \times 3 \times 256$\\

   Conv-2        & $2 \times 3$, Stride=(1, 1)  & $16 \times 2 \times 256$\\

   Conv-3        & $2 \times 3$, Stride=(1, 1)  & $32 \times 1 \times 256$\\
  \hline
   Flatten        &   ---  &     $8192$                  \\
   Linear-1        &   ---  &     $512$                    \\
   \textbf{Linear-2}        &   ---  &     $512$                    \\
   Linear-3        &   ---  &     $L+1$                    \\

  \hline
  \end{tabular}}
\end{table}

\renewcommand\arraystretch{1.3}
\begin{table}[t!]
  \centering
  \caption{Architecture of the CNN-Mask \cite{subramanian2022deep}.}
  \label{tab:cnn-mask}
  \scalebox{0.92}{\begin{tabular}{c c c}
  \hline
  Layer name &Structure & Output size \\
  \hline
   Input           & ---    &   $1 \times 4 \times 256$    \\

   Conv-1        & $2 \times 1$, Stride=(1, 1)  & $4 \times 3 \times 256$\\

   Conv-2        & $2 \times 3$, Stride=(1, 1)  & $16 \times 2 \times 256$\\

   Conv-3        & $2 \times 3$, Stride=(1, 1)  & $32 \times 1 \times 256$\\
  \hline
   Flatten        &   ---  &     $8192$                  \\
   Linear-1        &   ---  &     $512$                    \\
   \textbf{Mask}       &   ---  &     $512$                    \\
   Linear-3        &   ---  &     $L+1$                    \\

  \hline
  \end{tabular}}
\end{table}

\subsubsection{CNN-Mask}

Inspired by \cite{subramanian2022deep}, we designed a {mask layer} for both single and multiple speakers, implemented with Bi-directional Long Short-Term Memory (BiLSTM). Table~\ref{tab:cnn-mask} outlines the CNN architecture incorporating the mask layer. We replaced the original second dense layer in CNN-MLC with $B$ parallel BiLSTM layers, which results in the CNN-Mask backbone network. The $B$ BiLSTM layers take the sigmoid function as the activations. They are designed to learn $B$ ratio masks. See the following for the details.

We denote the embedding features produced from the first dense layer as $E \in \mathbb{R}^{T \times D}$, where $D$ is the embedding dimension, and $E$ comprises features from direct sounds, reverberations, and noises. We aim to implicitly isolate direct sound features, which is represented as ratio masks:
\begin{equation}\label{eq:sep_mask2}
      \{W_b\}_{b=1}^B = \mathrm{Sep}(E)
\end{equation}
where $W_b \in [0, 1]^{T \times D}$ represents the ratio mask for speaker $b$, and $\mathrm{Sep}(\cdot)$ denotes the {mask layer}. Consequently, the embedding feature of the direct sound of speaker $b$, denoted as $\boldsymbol{e}_b \in \mathbb{R}^D$, can be recovered by applying the mask $W_b$ to $E$ through element-wise multiplication:
\begin{equation}\label{eq:sep_mask3}
      \boldsymbol{e}_b = \frac{\sum_{t=1}^T W_b \times E}{\sum_{t=1}^T W_b}
\end{equation}
which is further processed through a dense layer to derive the predicted distribution $\boldsymbol{\hat \rho}_b$ for speaker $b$:
\begin{equation}\label{eq:sep_mask4}
      \boldsymbol{\hat \rho}_b = \mathrm{Dense}(\boldsymbol{e}_b)
\end{equation}
where $\mathrm{Dense}(\cdot)$ is composed of a linear layer with the softmax activation.

\subsection{Permutation ambiguity} \label{subsec:permutation}
Training the CNN-Mask backbone network when $B>1$ involves speaker separation, which encounters the permutation ambiguity problem. This subsection describes two ways to address the issue.

\subsubsection{Permutation invariant training}

Intuitively, this can be addressed using permutation invariant training (PIT) \cite{yu2017permutation}. We briefly outline PIT as follows:
\begin{equation}
\mathcal{L}_{\text {PIT }}=\min _{\boldsymbol{\psi} \in \Psi} \sum_{b=1}^B \mathcal{L}(\boldsymbol{\hat \rho}_b, \boldsymbol{\rho}_{\psi(b)})
\end{equation}
where $\boldsymbol{\rho}_b$ denotes the label distribution for speaker $b$, $\mathcal{L}$ stands for a loss function, $\Psi$ is a set encompassing all permutations of $B$ speakers, and $\boldsymbol{\psi}$ represents an individual permutation, with $\psi(b)$ indicating the $b$-th speaker in the permutation $\boldsymbol{\psi}$.

\subsubsection{Location-based training}

An alternative training method to address the permutation ambiguity problem in the multi-channel scenarios is the location-based training (LBT) \cite{taherian2022multi}. {LBT arranges $B$ speakers in the order of their DOAs.} For example, for a linear array with an azimuth range of $[0,180]^{\circ}$, suppose speaker $1$ locates at the $30^{\circ}$ angle, speaker $2$ at $90^{\circ}$, and speaker $3$ at $60^{\circ}$, then the speaker sequence can be $\{1,3,2\}$, which overcomes the speaker ambiguity problem.

In this paper, we apply LBT to the model training of CNN-Mask. Specifically, suppose the original label list is denoted as $\boldsymbol{P} = [\boldsymbol{\rho}_{b}]_{b=1}^B$. By sorting this list based on the DOA value, we obtain $\boldsymbol{P}^\mathrm{lbt} = [\boldsymbol{\rho}_{b}^\mathrm{lbt}]_{b=1}^B$. It allows us to compute the loss function as follows:
\begin{equation}
\mathcal{L}_{\text {LBT }}= \sum_{b=1}^B \mathcal{L}(\boldsymbol{\hat \rho}_b, \boldsymbol{\rho}_b^\mathrm{lbt})
\end{equation}
Training CNN-Mask with LBT not only significantly reduces the computational costs but also improves the performance.

\subsection{Label encoding} \label{subsec:uld}
Because the aforementioned CNN-based DOA estimation methods are formulated as classification problems where one-hot encoding is used to generate training labels, it suffers quantization errors inevitably. This paper proposed to apply unbiased label distribution (ULD) \cite{feng2023rethinking} as an alternative to one-hot, which completely eliminates quantization errors.

Here we apply ULD to the linear array as an example. In the full azimuth range of 180 degrees, the inter-class interval is
\begin{equation}
r = 180 / L
\end{equation}
We denote the ground-truth DOA of a speaker as $\theta$, then its corresponding class is $\gamma = \theta/r$, and its corresponding label distribution $\boldsymbol{\rho}$ is, then they have the following connection:
\begin{small}
\begin{equation}\label{eq:uld}
    \rho_{i} =
    \left\{\begin{array}{ll}
      1-\mathrm{deci}(\gamma),& \mbox{ if   } l = \mathrm{int}(\gamma) \\
      \mathrm{deci}(\gamma),& \mbox{ if   } l = \mathrm{int}(\gamma)+1 \\
      0,& \mbox{ otherwise}
    \end{array}\right.,\quad \forall l = 0,\ldots, L
\end{equation}
\end{small}
where $\mathrm{int}(\gamma)$ denotes the integer part of $\gamma$, and $\mathrm{deci}(\gamma)$ denotes the decimal part of $\gamma$.

\subsection{Weighted adjacent decoding} \label{subsec:wad}

In the test stage, we employ weighted adjacent decoding (WAD) \cite{feng2023rethinking} to project the estimated $\hat{\boldsymbol{\rho}}$ to an azimuth angle $\hat{\theta}$ without quantization errors, instead of the conventional one-best decoding. WAD is briefly described as follows. We identify the class with the highest probability as $\hat k = \arg \max_i \{\hat{\rho}_l\}_{l=0}^L$. WAD is defined as follows:
\begin{equation}\label{eq:wad3}
    \hat \theta = \frac{\sum_{l=\{\hat k-1, \hat k, \hat k+1\}} \hat \rho_l \times l \times  r}{\sum_{l=\{\hat k-1, \hat k, \hat k+1\}} \hat \rho_l}
\end{equation}
Note that in practice, we usually pad a zero element on each side of $\hat{\boldsymbol{\rho}}$, i.e. $\widehat{\boldsymbol{\rho}}= [0,\hat{\boldsymbol{\rho}},0]$, to accommodate the special case that $\hat{k}$ appears to be 0 or $L$.

\section{Node interaction} \label{sec:inter}
\subsection{Node selection}
Ad-hoc microphone arrays are distributed throughout a large acoustic scene. When a node is far from speech sources, the signal-to-noise ratio (SNR) of the signal collected by the node is low. Our preliminary studies show that (i) SNR strongly affects the accuracy of the DOA estimation at the node, and (ii) taking all nodes with the SNR varying in a large range results in large DOA estimation errors. Therefore, we need to select only the nodes with high SNRs for the 2D coordinate estimation.

As the first study of the node selection for the deep-learning-based speaker localization, this work follows the K-best node selection method in \cite{zhang2021deep}. Different from \cite{zhang2021deep} which needs an additional deep model for the SNR estimation, we conduct node selection based on the DOA estimations from all nodes, i.e. {$\{{\hat \rho}_{(n,b,l)}\}_{n=1,b=1,l=0}^{N,B,L}$}, from the CNN models.

Inspired by WAD \cite{feng2023rethinking}, we calculate the importance of each node by:
\begin{equation}\label{eq:wad_score}
  \delta_{n} = \frac{1}{B} \sum_{b=1}^B \sum_{l=\{\hat k-1, \hat k, \hat k+1\}} \hat \rho_{n,b,l}, \quad\forall n = 1,\ldots,N
\end{equation}
and pick the top $K$ nodes with the largest $\delta_n$, $\forall n = 1,\ldots,N$. Our preliminary study shows that the score in \eqref{eq:wad_score} has a strong positive correlation with SNR. To save the space of this paper, we omit the demonstration here.

\subsection{Triangulation for rough position estimation}


\begin{figure}[t]
  \centering
  \includegraphics[width=1.5in]{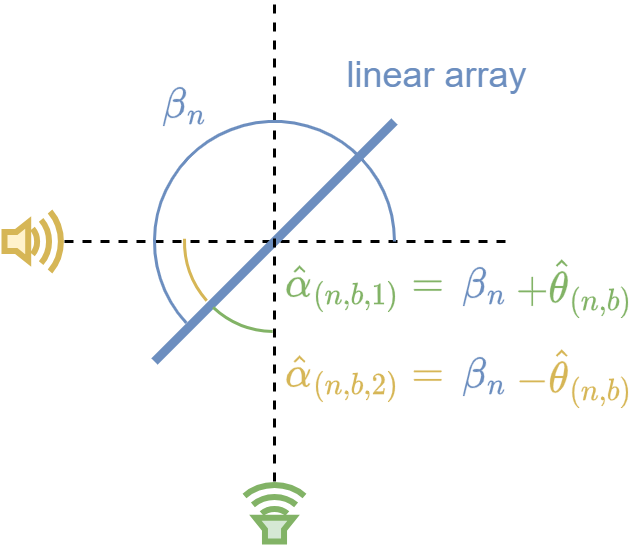}
  \caption{The Eq. \eqref{eq:alpha_nonlinear} and \eqref{eq:alpha_linear} can be interpreted physically as follows: For nonlinear arrays, only one bearing line $\hat{\alpha}_{(n,b,1)}$ will be emitted. However, for linear arrays, two bearing lines $\hat{\alpha}_{(n,b,1)}$ and $\hat{\alpha}_{(n,b,2)}$ will be emitted.}
  \label{fig2}
  \end{figure}

We use a triangulation method to get the 2D coordinates of speakers. It regards the estimated DOAs as bearing lines, and takes the cross between the bearing lines as the speaker positions. Specifically, we build a coordinate system in the room. Suppose the 2D coordinate of the $n$-th node is $\mathbf{l}_n=[x_n,\ y_n]$, and the self-angle of the node with respect to the room coordinate system is $\beta_{n}$. Suppose the ground-truth 2D coordinate of the $b$-th speaker is $\mathbf{c}_b=[{x}_b, \ {y}_b]$.

We denote the DOA estimation of the $b$-th speaker by the $n$-th node as $\hat{\theta}_{(n,b)}$. Note that, the proposed algorithm does not have to know which speaker the estimation $\hat{\theta}_{(n,b)}$ belongs to; the index $b$ is only used to remind that the proposed algorithm is suitable for multi-speaker localization. The 2D speaker localization problem is formulated as estimating $\mathbf{c}_b$ given $\{\mathbf{l}_n\}_{n=1}^{K}$, $\{\beta_{n}\}_{n=1}^{K}$, and $\{\hat{\theta}_{(n,b)}\}_{n=1,b=1}^{K,B}$. We denote $\hat{\mathbf{c}}_b = [\hat{x}_b, \ \hat{y}_b]$ as an estimation of $\mathbf{c}_b$, and define $\hat{\alpha}_{(n,b)}$ as an angle between $\mathbf{l}_n$ and $\hat{\mathbf{c}}_b$ in the coordinate system:
\begin{equation}
  \tan{\hat{\alpha}_{(n,b)}=\frac{\hat{y}_b-y_n}{\hat{x}_b-x_n}}
\end{equation}

Starting from the scenario where each node is a nonlinear array, we can obtain $B$ bearing lines $\hat{\alpha}_{(n,b)}$ from a node by:
\begin{equation}\label{eq:alpha_nonlinear}
  \hat{\alpha}_{(n,b)} = (\beta_{n} + \hat{\theta}_{(n,b)}) \% 360
\end{equation}
where the symbol ``\%'' denotes the modulo operator, aiming to ensure $\hat{\alpha}_{(n,b)} \in [0, 360)$. As shown in Fig.~\ref{fig2}, the physical meaning of ~\eqref{eq:alpha_nonlinear} is to rotate $\hat{\theta}_{(n,b)}$ degrees anticlockwise based on the self-angle $\beta_{n}$ of the array.

\begin{figure}[t]
\centering
\includegraphics[width=1.5in]{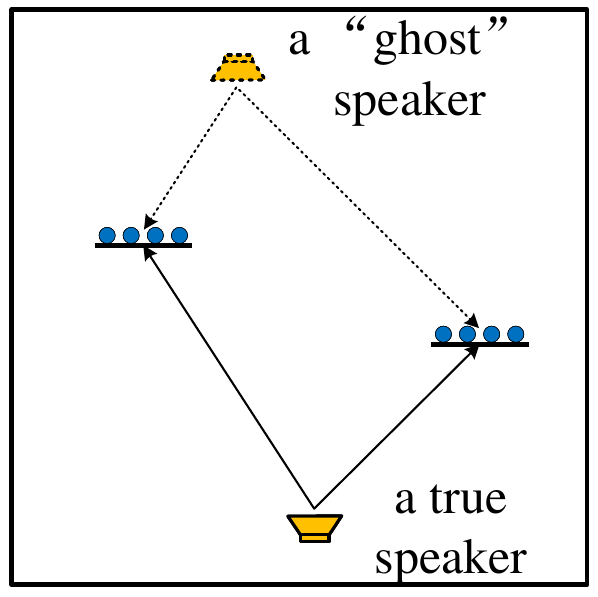}
\caption{The ``ghost'' speaker problem caused by linear arrays.}
\label{fig6}
\end{figure}

 On the other side, a linear array emits two bearing lines for a single speaker, as illustrated in Fig.~\ref{fig6}. Therefore, the situation where each node is a linear array is more complex than that with a nonlinear array. We can obtain $2B$ bearing lines $\{\hat{\alpha}_{(n,b,c)}\}_{n=1,b=1,c=1}^{K,B,2}$ from each node by:
\begin{equation}\label{eq:alpha_linear}
  \begin{array}{ll}
  \hat{\alpha}_{(n,b,1)} = (\beta_{n} + \hat{\theta}_{(n,b)}) \% 360 \\
  \hat{\alpha}_{(n,b,2)} = (\beta_{n} - \hat{\theta}_{(n,b)}) \% 360
\end{array}
\end{equation}

As shown in Fig.~\ref{fig2}, the difference between \eqref{eq:alpha_linear} and \eqref{eq:alpha_nonlinear} is that there is an additional clockwise bearing line.


Then, the triangulation method can get a speaker position from any two bearing lines of a pair of nodes at a time. Suppose we use the $n_1$-th and $n_2$-th nodes where we know the 2D coordinates of the two nodes $(x_{n_1},y_{n_1})$ and $(x_{n_2},y_{n_2})$ and their estimated DOA angles $\hat{\alpha}_{(n_1,b)}$ and $\hat{\alpha}_{(n_2,b)}$, it is easy to obtain an estimated 2D speaker position $[\hat{x}_b, \hat{y}_b]$ of the $b$-th speaker as shown in Fig. \ref{fig1}:
  \begin{equation}
  \setlength{\arraycolsep}{0.5pt}
  \renewcommand{\arraystretch}{2}
  \begin{array}{l}
  \hat{x}_b=\frac{x_{n_1}\tan\hat{\alpha}_{(n_1,b)}-x_{n_2} \tan\hat{\alpha}_{(n_2,b)}+y_{n_2}-y_{n_1}}{\tan\hat{\alpha}_{(n_1,b)}-\tan\hat{\alpha}_{(n_2,b)}},\\
  \hat{y}_b=\frac{\tan\hat{\alpha}_{(n_1,b)}\tan\hat{\alpha}_{(n_2,b)}(x_{n_1}-x_{n_2})+y_{n_2} \tan\hat{\alpha}_{(n_1,b)}-y_{n_1} \tan\hat{\alpha}_{(n_2,b)}}{\tan\hat{\alpha}_{(n_1,b)}-\tan\hat{\alpha}_{(n_2,b)}}
  \end{array}
  \end{equation}
However, because any two bearing lines can get a speaker position, then the estimated speaker positions from the whole system is exponentially larger than the number of ad-hoc nodes. In other words, the triangulation does not address the multi-speaker localization problem. This difficulty can be solved by the following clustering strategy.

Note that linear arrays have a ghost speaker problem. As illustrated in Fig.~\ref{fig6}, because it is well known that a linear array yields two DOA estimations for a single speaker, four bearing lines from a pair of linear arrays result in two speaker positions, one of which is a ``ghost'' speaker.

\subsection{Clustering for precise position estimation}

We employ kernel mean-shift clustering to integrate the rough 2D estimates produced from the triangulation module into accurate speaker positions. Mean-shift clustering is a nonparametric clustering algorithm. It first finds the area of the data points with the highest intensity and then gets the cluster centers by a mean shift procedure. Here we choose the Gaussian kernel mean-shift clustering \cite{wu2007mean}, which takes $B$ cluster centers $\{\hat{\mathbf{c}}_{b}\}_{b=1}^B$ as the accurately estimated speaker positions. The advantage of the clustering strategy is that a biased DOA estimation of a single node has little negative effect on the global speaker localization. Based on this advantage, if the ``ghost'' speaker positions caused by the triangulation module are not too many, they may not have strong negative effects to $\{\hat{\mathbf{c}}_{b}\}_{b=1}^B$.

\section{{Datasets}} \label{sec:data}
\subsection{Simulated data} \label{subsec:data_simu}

We used the  `train-clean-360', `dev-clean', and `test-clean' subsets of the LibriSpeech corpus \cite{panayotov2015librispeech} as the clean speech sources of the training, validation, and test sets of the simulated data, respectively. Additive ambient noise was added into the simulated data. The noise source  \cite{tan2021speech} contains 126 hours of various noise types. The noise segments in the training, validation, and testing sets did not overlap.
 
For the training and validation sets, we only need to generate simulation data for a single fixed array, either circular or linear, since that they will only be used to the training of the CNN-based DOA estimation. Specifically, each utterance was 2 seconds long. For each individual utterance, we generated a room. The length and width of the room were randomly generated from a range of $[4, 10]$ meters. The height of the room was randomly generated from $[3, 4]$ meters. A single microphone array and one to two speakers were randomly placed in the room. The heights of both the microphone array and the speakers were set to $1.3$ meters. Each circular array or linear array contains 4 microphones with an aperture of 8 cm. The self-angle of the microphone array was randomly chosen. We used Pyroomacoustics \cite{scheibler2018pyroomacoustics} to generate the room impulse response. The reverberation time T$_{60}$ was randomly chosen from a range of $[0.2, 1.0]$ seconds. The SNR was randomly drawn from a range of $[0, 20]$ dB. Each training set comprises 24,000 utterances. Each validation set consists of 1,200 utterances.
  
For the test sets, we need to generate simulated data for ad-hoc microphone arrays, whose ad-hoc nodes are either circular arrays or linear arrays. Specifically, for each randomly generated room, we repeated the procedure of constructing the training data, except that (i) we randomly placed 10 ad-hoc nodes in the room and (ii) we placed $B$ speakers in the room with $B=\{1,2\}$. We added diffuse noise with an SNR level randomly selected from $[10, 20, 30]$ dB. The SNR was calculated as an energy ratio of the average direct sound of all microphone channels to the diffuse noise. Note that, due to the potential large difference in distances between the nodes and speakers, the SNR at the nodes could vary in a wide range. Each test set consists of 1,200 utterances. To study the effects of different types of microphone arrays on performance, for each randomly generated test room, we applied exactly the same environmental setting (including the speech source, room environment, speaker positions, microphone node positions and self-angles) to both circular-array-based ad-hoc nodes and linear-array-based ad-hoc nodes.

To summarize, for either the circular- or the linear-array-based scenario, we prepared 2 training set, 2 validation set, 3 single-speaker test sets with SNR levels of $[10, 20, 30]$ dB respectively, and 3 two-speaker test sets with a similar SNR setting.

\subsection{Real-world data}
\begin{figure*}[t]
  \begin{center}
  \subfigure[Configuration 1, office]{
  \includegraphics[width=1.7in]{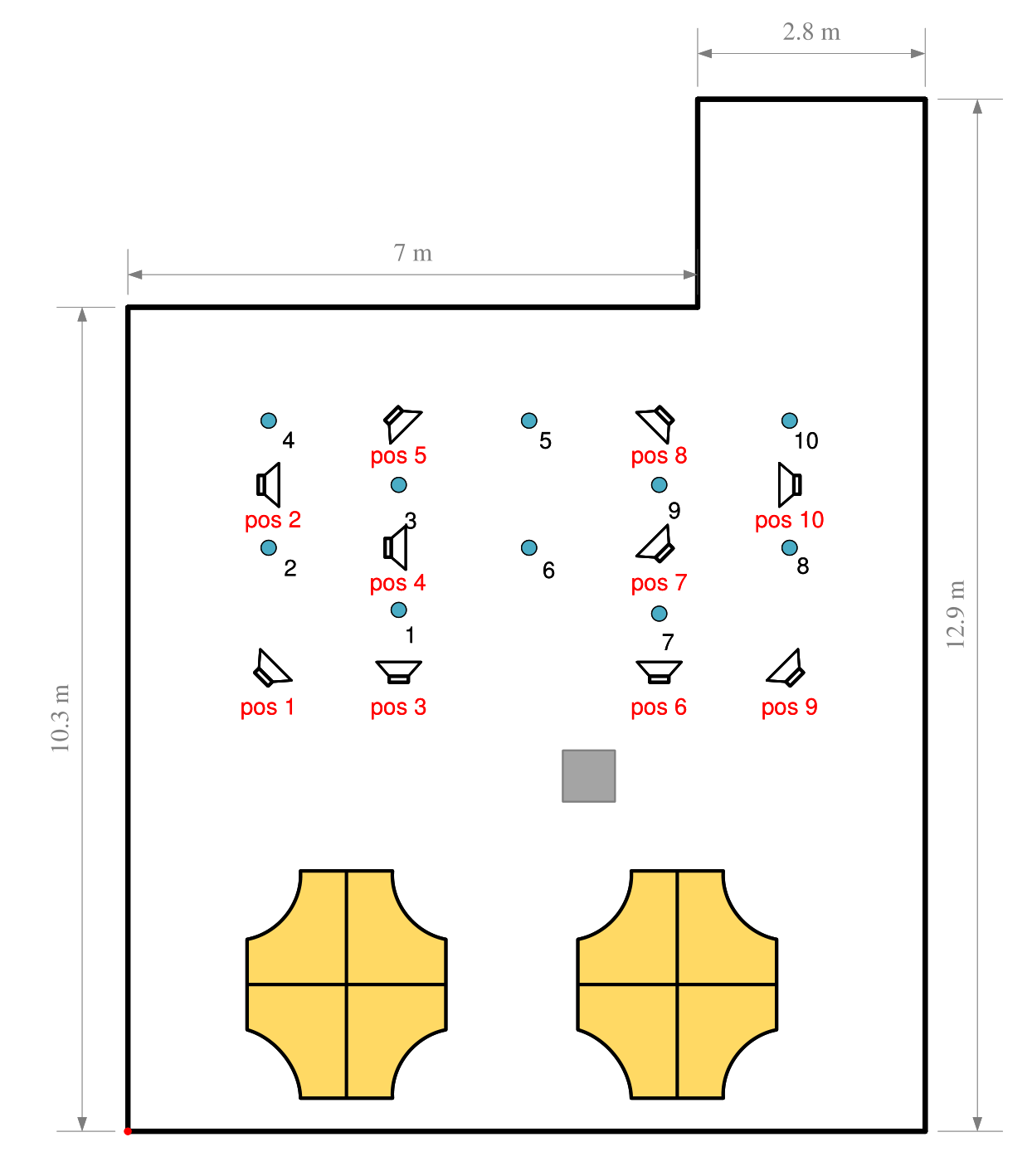}
  }
  \subfigure[Configuration 2, office]{
  \includegraphics[width=1.7in]{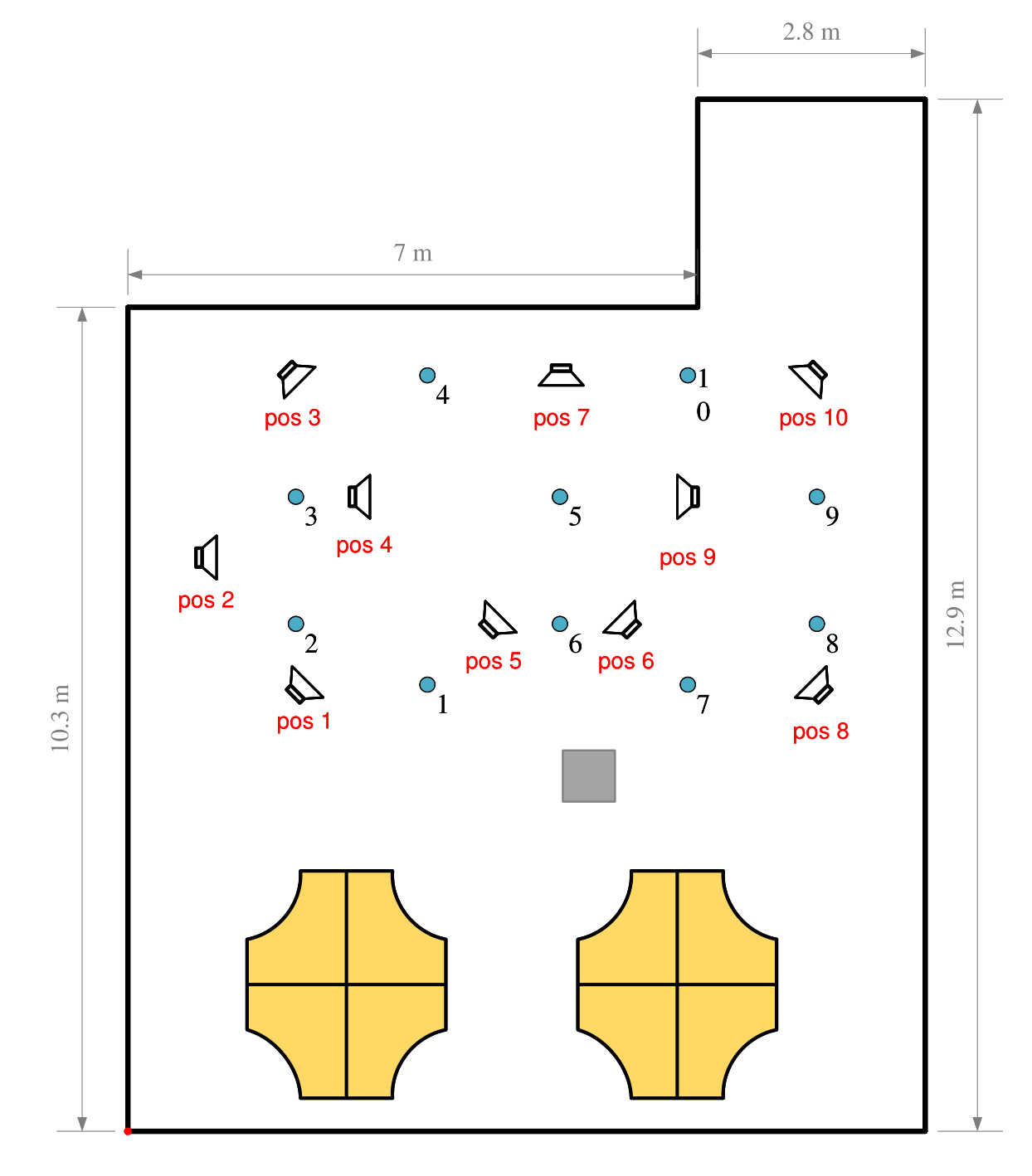}
  }
  \subfigure[Configuration 1, conference]{
  \includegraphics[width=1.5in]{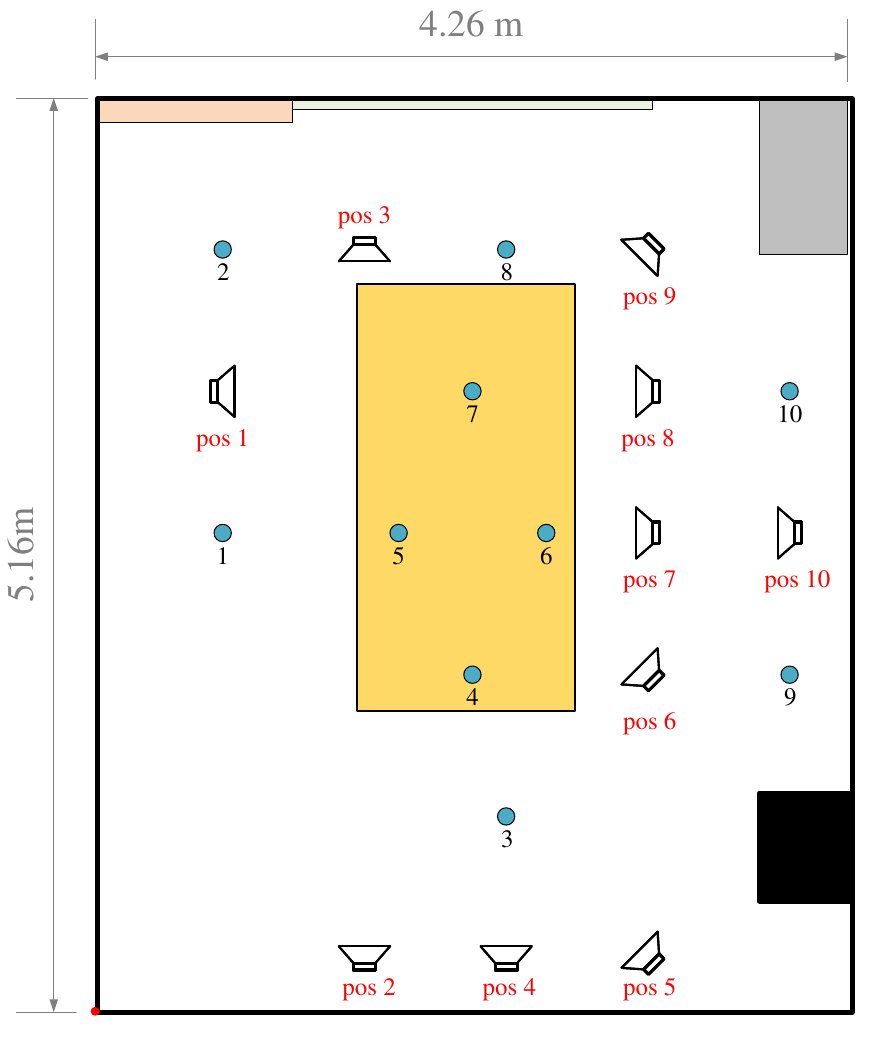}
  }
  \subfigure[Configuration 2, conference]{
  \includegraphics[width=1.5in]{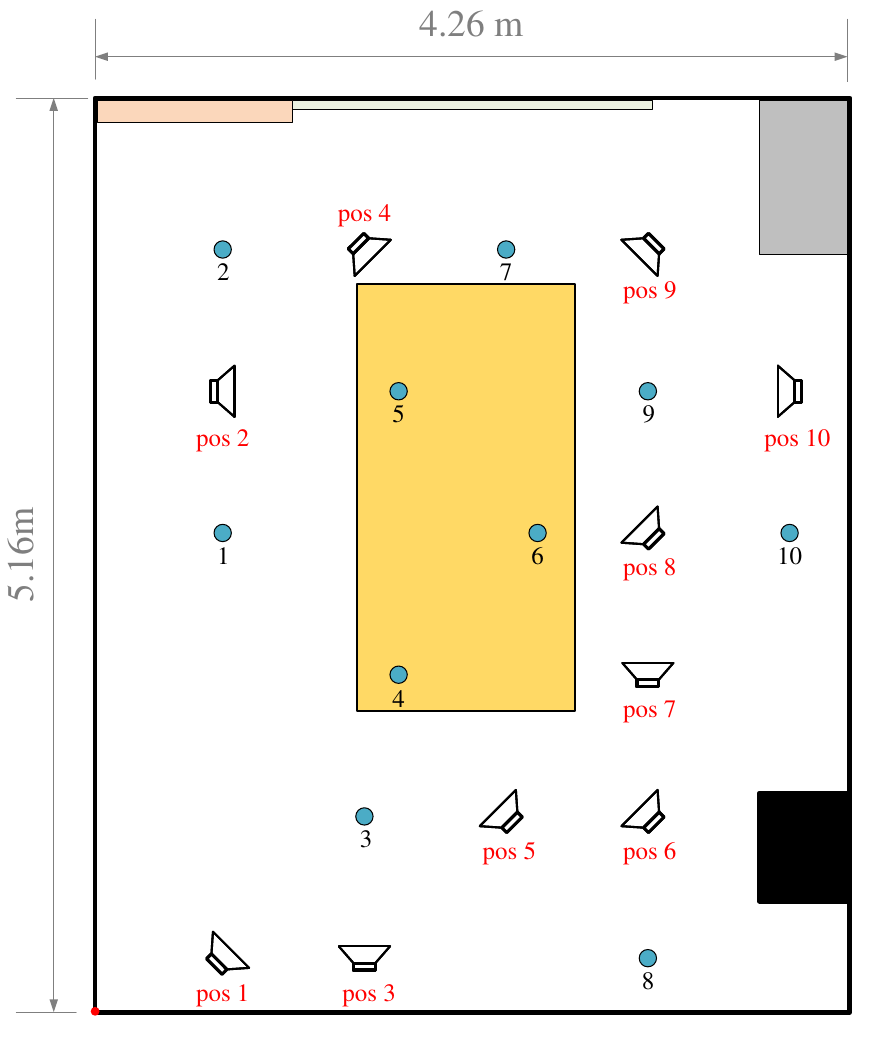}
  }
  \caption{Recording environment and settings of the two rooms of Libri-adhoc-nodes10. The blue dots represent the positions of the ad-hoc nodes. The loudspeaker icons represent the positions and orientations of the speakers. Each configuration has two sub configurations. The difference between the sub configurations lies in the different self-rotation angles of the sub arrays. }\label{fig:room}
  \end{center}
\end{figure*}

We recorded a real-world dataset named Libri-adhoc-node10. It contains a conference room and an office room. Each room has 10 ad-hoc nodes and a loudspeaker. Each node contains a 4-channel linear array with an aperture of 8cm. Fig. \ref{fig:room} shows the recording environment of the two rooms. The size of the office room is approximately $9.8 \times 10.3 \times 4.2$m with T$_{60}\approx 1.39$s. The size of the conference room is approximately $4.26 \times 5.16 \times 3.16$m with T$_{60}\approx 1.06$s. It records the `test-clean' subset of the LibriSpeech data replayed by the loudspeaker in the rooms, which contains 20 male speakers and 20 female speakers. The ad-hoc nodes and the loudspeaker have the same height of $1.3$m. The ambient noise of the recording environments can be ignored. The detailed description of the data and its open source, which includes the speaker ID and positions, microphone node positions, self-rotation angles, etc, will be released in \href{https://github.com/Liu-sp/Libri-adhoc-nodes10}{https://github.com/Liu-sp/Libri-adhoc-nodes10}. 

The above recorded dataset is a single-source dataset. We randomly selected 1200 2-second segments for testing. In addition, we synthesized a two-source dataset with 1200 segments by summing the recorded samples from different speaker positions.

\section{Empirical evaluation} \label{sec:exp}

\subsection{Parameter setting}
The sampling rate of the speech recordings was 16 kHz. {The number of STFT points was set to 512. The window length is 512, with a hop length of 256.} We excluded the zero frequency sub-band of STFT and extracted the phase spectrogram of the STFT feature from each of the four microphones of a node, resulting in a $4\times256$ phase map, which was used as input of the CNN-based DOA estimation. {The inter-class interval for classification was set to $5^{\circ}$, so that the number of the classes $L$ equals to $72$ for the circular array and $36$ for the linear array.} The bandwidth of the mean-shift clustering was set to 0.4, which is the default setting.

For training the CNN backbone networks, we used the AdamW \cite{loshchilov2019decoupled} optimizer with a batch size of 64. We trained the models by 30 epochs, with an initial learning rate of $10^{-3}$. If the validation loss did not decrease for 3 consecutive epochs, we reduced the initial learning rate to $10^{-4}$. The model with the minimum localization error on the validation set was chosen for evaluation. The best models on the simulated validation set were used for both simulated test data and real-world data. For the CNN-MLC backbone network, the output layer used sigmoid activation, and took binary cross entropy as the loss function. For the CNN-Mask network, the softmax activation and cross entropy were used.

\subsection{Evaluation metrics}
We used mean angle error (MAE) to evaluate the DOA estimation at each ad-hoc node:
\begin{equation}\label{eq:mae}
  \mathrm{MAE}= \mathrm{min}(|\hat \theta - \theta|, 360^{\circ}-|\hat \theta - \theta|)
\end{equation}

We used mean distance error (MDE) between the ground-truth speaker position $(x, y)$ and the predicted speaker position $(\hat{x}, \hat{y})$ to evaluate the performance of the comparison methods with ad-hoc microphone arrays:
\begin{equation}\label{eq:mde}
  \mathrm{MDE}= \sqrt{(\hat{x}-x)^{2}+(\hat{y}-y)^{2}}
\end{equation}

\subsection{Comparison methods}

\begin{table}[t]
  \centering
  \caption{Description of the comparison CNN-based methods, where ``\checkmark'' implies that the option is used.}
  \scalebox{0.74}{\begin{tabular}{cc|cc|cc|cc}
    \toprule
    Scenario & Method & MLC   & Mask  & One-hot & ULD   & PIT   & LBT \\
    \midrule
    \multirow{3}[2]{*}{Single source} & CNN-MLC & \checkmark &       & \checkmark &       &       &  \\
          & CNN-Mask &       & \checkmark & \checkmark &       &       &  \\
          & CNN-ULD &        & \checkmark &       & \checkmark &       &  \\
    \midrule
    \multirow{4}[2]{*}{Multiple source} & CNN-MLC & \checkmark &       & \checkmark &       &       &  \\
          & CNN-PIT &       & \checkmark & \checkmark &       & \checkmark &  \\
          & CNN-LBT &       & \checkmark & \checkmark &       &       & \checkmark \\
          & CNN-ULD &       & \checkmark &       & \checkmark &       & \checkmark \\
    \bottomrule
    \end{tabular}}
  \label{tab:methods}%
\end{table}%

We created three conventional algorithms using ad-hoc microphone arrays as the comparison baselines. They shared the same stage-wise 2D localization framework as the proposed method, except that the CNN-based DOA estimation was replaced by a conventional broadband DOA estimation method, such as MUSIC \cite{schmidt1986multiple}, SRP-PHAT \cite{dibiase2000high}, or NormMUSIC \cite{salvati2014incoherent}. The resolution was set to $5^{\circ}$. {The CNN-based methods for comparison are summarized in Table~\ref{tab:methods}, which are a set of diverse combinations of the technologies discussed in this paper.}

Note that the difference between the comparison methods is the DOA estimation module. Therefore, we used the name of the DOA estimation algorithm to represent the 2D speaker localization algorithm with ad-hoc microphone arrays without confusion. \textbf{In the following experiments, if we use MAE as the evaluation metric, it means that we are comparing the DOA estimation modules; if we use MDE as the evaluation metric, it means that we are comparing the 2D speaker localization algorithms.}

\subsection{Results on simulated data}\label{subsec:simu}

\subsubsection{Single-source localization}
\begin{table}[t]
  \centering
  \caption{MAE (in degrees) of the comparison DOA estimation methods on the simulated single-source data.}
  \scalebox{0.8}{\begin{tabular}{cccc|ccc}
    \toprule
    \multirow{2}[4]{*}{Method} & \multicolumn{3}{c}{Circular array} & \multicolumn{3}{c}{Linear array} \\
\cmidrule{2-7}          & 10dB  & 20dB  & 30dB  & 10dB  & 20dB  & 30dB \\
    \midrule
    MUSIC & 15.551 & 13.720 & 13.510 & 13.866 & 12.848 & 12.776 \\
    SRP-PHAT & 8.580 & 6.403 & 6.012 & 8.140 & 5.954 & 5.538 \\
    NormMUSIC & 9.546 & 7.849 & 7.795 & 8.624 & 7.103 & 7.038 \\
    \midrule
    CNN-MLC & 4.521 & 3.385 & 3.186 & 4.675 & 3.653 & 3.544 \\
    CNN-Mask & 4.950 & 3.624 & 3.163 & 4.456 & 3.347 & 3.191 \\
    {CNN-ULD} & \textbf{3.018} & \textbf{2.129} & \textbf{1.826} & \textbf{3.663} & \textbf{2.585} & \textbf{2.427} \\
    \bottomrule
    \end{tabular}}
  \label{tab:mae-1}%
\end{table}%

\begin{table}[t]
  \centering
  \caption{MDE (in meters) of the comparison 2D speaker localization methods on the simulated single-source data.}
  \scalebox{0.85}{\begin{tabular}{cccc|ccc}
    \toprule
    \multirow{2}[4]{*}{Method} & \multicolumn{3}{c}{Circular array} & \multicolumn{3}{c}{Linear array} \\
\cmidrule{2-7}          & 10dB  & 20dB  & 30dB  & 10dB  & 20dB  & 30dB \\
    \midrule
    MUSIC & 0.612 & 0.537 & 0.522 & 2.187 & 1.947 & 1.977 \\
    SRP-PHAT & 0.483 & 0.375 & 0.368 & 1.397 & 1.049 & 0.905 \\
    NormMUSIC & 0.520 & 0.442 & 0.420 & 1.605 & 1.218 & 1.194 \\
    \midrule
    CNN-MLC & 0.165 & 0.136 & 0.128 & 0.480 & 0.401 & 0.387 \\
    CNN-Mask & 0.154 & 0.130 & 0.113 & 0.397 & 0.336 & 0.332 \\
    {CNN-ULD} & \textbf{0.103} & \textbf{0.093} & \textbf{0.086} & \textbf{0.341} & \textbf{0.231} & \textbf{0.233} \\
    \bottomrule
    \end{tabular}}
  \label{tab:mde-1}%
\end{table}%

\begin{table}[t]
  \centering
  \caption{MDE (in meters) of the CNN-ULD-based system on the simulated single-source data with respect to the bandwidth of the mean shift clustering.}
  \scalebox{0.8}{\begin{tabular}{cccc|ccc}
    \toprule
    \multirow{2}[4]{*}{Bandwidth} & \multicolumn{3}{c}{Circular array} & \multicolumn{3}{c}{Linear array} \\
\cmidrule{2-7}          & 10dB  & 20dB  & 30dB  & 10dB  & 20dB  & 30dB \\
    \midrule
    0.1   & 0.141 & 0.109 & 0.096 & 0.400 & 0.340 & 0.337 \\
    0.2   & 0.114 & 0.097 & 0.087 & 0.305 & 0.253 & 0.256 \\
    0.3   & 0.104 & \textbf{0.091} & 0.087 & \textbf{0.299} & 0.240 & 0.239 \\
    {0.4} & \textbf{0.103} & 0.093 & \textbf{0.086} & 0.341 & \textbf{0.231} & \textbf{0.233} \\
    0.5   & 0.108 & 0.092 & 0.089 & 0.345 & 0.271 & 0.281 \\
    \bottomrule
    \end{tabular}}
  \label{tab:mde-1-ms}%
\end{table}%

\begin{table}[t]
  \centering
  \caption{MDE (in meters) of the CNN-ULD-based system on the simulated single-source data with respect to the number of the selected ad-hoc nodes.}
  \scalebox{0.8}{\begin{tabular}{cccc|ccc}
    \toprule
    \multirow{2}[4]{*}{Nodes} & \multicolumn{3}{c}{Circular array} & \multicolumn{3}{c}{Linear array} \\
\cmidrule{2-7}          & 10dB  & 20dB  & 30dB  & 10dB  & 20dB  & 30dB \\
    \midrule
    6     & 0.111 & 0.097 & 0.097 & 0.360 & 0.324 & 0.337 \\
    7     & \textbf{0.098} & 0.098 & 0.093 & 0.326 & 0.276 & 0.265 \\
    8     & 0.099 & \textbf{0.089} & \textbf{0.082} & 0.310 & 0.249 & 0.248 \\
    9     & 0.101 & 0.090 & 0.085 & \textbf{0.287} & 0.255 & 0.237 \\
    10    & 0.103 & 0.093 & 0.086 & 0.341 & \textbf{0.231} & \textbf{0.233} \\
    \bottomrule
    \end{tabular}}
  \label{tab:mde-1-node}%
\end{table}%

Table \ref{tab:mae-1} lists the MAE scores of the comparison DOA estimation methods on the simulated single-source data. From the table, we observe that the MAEs of the comparison methods with either the circular array or the linear array are quite similar. The CNN-based methods consistently outperform the conventional baselines. CNN-Mask outperforms CNN-MLC in most cases, which indicates the effectiveness of the mask layers. Furthermore, the proposed CNN-ULD can mitigate quantization errors, which helps it consistently achieve the lowest MAE.

Table \ref{tab:mde-1} lists the MDE scores of the comparison 2D speaker localization methods with ad-hoc microphone arrays on the simulated single-source data. Comparing Table \ref{tab:mae-1} with Table \ref{tab:mde-1}, it is evident that, under the same sub-array type, the accuracies of the DOA estimation and the 2D speaker localization are generally positively correlated, and CNN-ULD performs the best in both cases.

One difference between Table \ref{tab:mae-1} and Table \ref{tab:mde-1} is that, when the circular array is used, the proposed framework for the 2D speaker localization is more robust against noise than its DOA estimation module. We take the results of CNN-ULD with the circular array as an example, when SNR decreases from 20dB to 10dB, its MAE increases relatively by 41.76\%, while its MDE only slightly increases by 10.75\%.

However, for the linear array, the ghost speakers introduce additional interference into the 2D speaker localization, which leads to higher MDE than the circular array. However, the results are still promising, indicating that mean-shift clustering exhibits robustness against ghost speakers to some extent.

We fine-tuned the bandwidth parameter of the mean-shift clustering in the CNN-ULD-based ad-hoc microphone array, where the node selection module was disabled. From Table \ref{tab:mde-1-ms}, it is evident that the circular array is insensitive to the bandwidth variation due to the absence of the interference from ghost speakers. The results with the linear array are also not highly sensitive to the bandwidth values: Selecting a value between 0.2 and 0.4 yields comparable outcomes.

We studied the node selection strategy by selecting different number of nodes. From the results in Table \ref{tab:mde-1-node}, we see that, the MDE of the CNN-ULD system with the circular array did not vary much with respect to the number of nodes. However, selecting limited number of nodes significantly reduces the computational complexity. In contrast, the CNN-ULD system with the linear array benefits more when the number of nodes was set large, especially in the high SNR scenarios. This phenomenon can be explained as that, if a large number of nodes were selected, then the interference of the ghost speakers on the mean-shift clustering may be reduced.

\subsubsection{Multi-source localization}

\begin{table}[t]
  \centering
  \caption{MAE (in degrees) of the comparison methods on the simulated two-source data.}
  \scalebox{0.8}{\begin{tabular}{cccc|ccc}
    \toprule
    \multirow{2}[4]{*}{Method} & \multicolumn{3}{c}{Circular array} & \multicolumn{3}{c}{Linear array} \\
\cmidrule{2-7}          & 10dB  & 20dB  & 30dB  & 10dB  & 20dB  & 30dB \\
    \midrule
    MUSIC & 45.648 & 44.964 & 44.915 & 29.838 & 29.822 & 29.819 \\
    SRP-PHAT & 34.643 & 33.263 & 32.983 & 23.310 & 21.606 & 21.293 \\
    NormMUSIC & 39.395 & 38.031 & 37.841 & 27.787 & 26.770 & 26.613 \\
    \midrule
    CNN-MLC & 28.691 & 27.838 & 27.369 & 20.117 & 18.598 & 18.533 \\
    CNN-PIT & 22.254 & 20.688 & 20.262 & 16.473 & 15.105 & 14.886 \\
    CNN-LBT & 18.698 & 16.718 & 16.234 & 13.258 & 11.660 & 11.319 \\
    {CNN-ULD} & \textbf{17.358} & \textbf{15.405} & \textbf{15.015} & \textbf{11.597} & \textbf{10.164} & \textbf{9.915} \\
    \bottomrule
    \end{tabular}}
  \label{tab:mae-2}%
\end{table}%

\begin{table}[t]
  \centering
  \caption{MDE (in meters) of the comparison methods on the simulated two-source data.}
  \scalebox{0.85}{\begin{tabular}{cccc|ccc}
    \toprule
    \multirow{2}[4]{*}{Method} & \multicolumn{3}{c}{Circular array} & \multicolumn{3}{c}{Linear array} \\
\cmidrule{2-7}          & 10dB  & 20dB  & 30dB  & 10dB  & 20dB  & 30dB \\
    \midrule
    MUSIC & 2.013 & 2.012 & 2.001 & 2.743 & 2.762 & 2.753 \\
    SRP-PHAT & 1.851 & 1.843 & 1.837 & 2.368 & 2.337 & 2.344 \\
    NormMUSIC & 1.869 & 1.851 & 1.857 & 2.492 & 2.413 & 2.405 \\
    \midrule
    CNN-MLC & 1.271 & 1.236 & 1.218 & 1.904 & 1.840 & 1.828 \\
    CNN-PIT & 1.129 & 1.132 & 1.080 & 1.873 & 1.818 & 1.809 \\
    CNN-LBT & 1.145 & 1.064 & 1.053 & 1.960 & 1.884 & 1.878 \\
    \textbf{CNN-ULD} & \textbf{0.987} & \textbf{0.890} & \textbf{0.878} & \textbf{1.840} & \textbf{1.814} & \textbf{1.799} \\
    \bottomrule
    \end{tabular}}
  \label{tab:mde-2}%
\end{table}%

Table \ref{tab:mae-2} lists the MAE results of the comparison DOA estimation algorithms on the simulated two-source data.
From the table, we see that the performance of all DOA algorithms in the multi-speaker localization scenarios drops compared to that in the single-speaker scenarios. However, the CNN-based methods still outperform conventional methods. CNN-ULD performs the best in the CNN-based methods, while CNN-LBT outperforms CNN-PIT. Finally, the DOA methods with the circular array perform worse than those with the linear arrays.

Table \ref{tab:mde-2} lists the MDE results of the comparison 2D speaker localization systems on the simulated two-source data.
From the table, we see that, although the linear arrays provide more accurate DOA estimates than the circular arrays as shown in Table \ref{tab:mae-2}, the MDE performance of the 2D speaker localization systems with the linear arrays are significantly worse than those with the circular arrays. This is mainly caused by that the ghost speakers introduced by the linear arrays are too many. We will discuss this issue in detail in Section \ref{sec:dis}.

\subsection{Results on real-world data}\label{subsec:real}

\begin{table}[t]
  \centering
  \caption{MDE (in meters) of the comparison 2D speaker localization methods on the real-world data. Note that in the single-source scenario, CNN-LBT degrades into CNN-Mask.}
  \scalebox{0.85}{\begin{tabular}{ccc|cc}
    \toprule
    \multirow{2}[4]{*}{Method} & \multicolumn{2}{c}{Conference} & \multicolumn{2}{c}{Office} \\
\cmidrule{2-5}          & 1 speaker & 2 speakers & 1 speakers & 2 speakers \\
    \midrule
    MUSIC & 1.712 & 1.654 & 3.606 & 3.816 \\
    SRP-PHAT & 1.518 & 1.299 & 3.768 & 3.821 \\
    NormMUSIC & 1.036 & 1.224 & 3.473 & 3.736 \\
\cmidrule{1-5}    CNN-MLC & 0.393 & 0.851 & 0.496 & 2.229 \\
    CNN-PIT & ---     & 0.842 & ---     & 2.234 \\
    CNN-LBT & 0.329 & 0.837 & 0.419 & 2.079 \\
    CNN-ULD & \textbf{0.229} & \textbf{0.837} & \textbf{0.260} & \textbf{1.958} \\
    \bottomrule
    \end{tabular}}
  \label{tab:mde-r}%
\end{table}%

Table \ref{tab:mde-r} lists the MDE results of the comparison 2D speaker localization systems with the linear arrays on the real-world data. From the table, we see that, although the CNN-based methods were only trained on the simulated data, they generalize well on the real-world data, and consistently outperform the conventional methods.
Due to the strong interference from ghost speakers, the MDE produced from conventional methods seem to be too large, which indicates that they perform random guess to the speaker positions. Therefore, our focus is on the CNN-based methods. (i) For the single-source localization, the MDE is controlled to a sufficiently low level. Comparing Tables \ref{tab:mde-r} and \ref{tab:mde-1}, we see that the MDE on the real-world data is close to that on the simulated data with 30dB, since that the real-world data is nearly ambient-noise-free. (ii) For the multi-source localization, all CNN-based methods show significant performance degradation due to the strong interference made by the ghost speakers. This negative effect will be further discussed in Section~\ref{sec:dis}.

\section{Discussions}\label{sec:dis}

This section systematically studies the effects of the DOA estimation errors and ghost speakers on the performance of the 2D speaker localization.

\subsection{Theoretical analysis of the negative effect of ghost speakers}

For nonlinear arrays, such as the circular array, they estimate a single DOA for a speaker, which produces $B$ DOA estimates for a $B$ speaker localization problem. Therefore, any two ad-hoc nodes with the circular arrays can have at most $B^2$ intersections, among which only $B$ points are supposed to be close to the ground-truth 2D positions of the speakers. Eventually, the two nodes generate $(B^2-B)$ ghost speakers. One special case is that, when $B=1$, the two nodes do not produce ghost speakers at all.

For the linear array, it estimates two DOAs for a speaker. Therefore, any two ad-hoc nodes with the linear array produce $4B^2-B$ ghost speakers, which is far more than the nonlinear arrays. Particularly, even if $B=1$, the two nodes still produce 3 ghost speaker at most.

The above analysis explains the phenomena that (i) comparing Tables \ref{tab:mae-1} and \ref{tab:mde-1} in the single-source scenario, although the the DOA estimation algorithms with the circular array slightly outperform their counterparts with the linear array, their 2D sound source localization systems are significantly better than those with the linear arrays; and moreover (ii) comparing Tables \ref{tab:mae-2} and \ref{tab:mde-2} in the two-source scenario, even if the the DOA estimation algorithms with the circular array are worse than their counterparts with the linear array, their 2D sound source localization systems are better than those with the linear arrays.

\subsection{Ablation study on ghost speakers and DOA estimation errors}

It is known that the MDE of the final 2D sound source localization is mainly affected by two factors: the MAE of the DOA estimation module, and the number of ghost speakers caused by the triangulation module. The former introduces the DOA estimation errors, while the latter further introduces ghost speaker interference. In this subsection, we designed a controllable experiment to study the effects of the two interference factors on performance quantitatively with respect to different MAE levels.

\subsubsection{Visualization}

We generated a room of $8 \times 8$ square meters. For the single-source scenario, one speaker was placed at $(2, 4)$. For the two-source scenario, two speakers were placed at $(2, 4)$ and $(6, 4)$ respectively. We randomly placed 10 ad-hoc nodes. The nodes were either circular arrays or linear arrays.
For each node, we further added random perturbation to the ground-truth DOA, so as to control the expectation of the MAE of the node to be at levels of $[0 , 1, 5, 15]$ degrees respectively.

\begin{figure*}
  \centering
  \subfigure[1 speaker, circular arrays]{
    \includegraphics[width=0.92\textwidth]{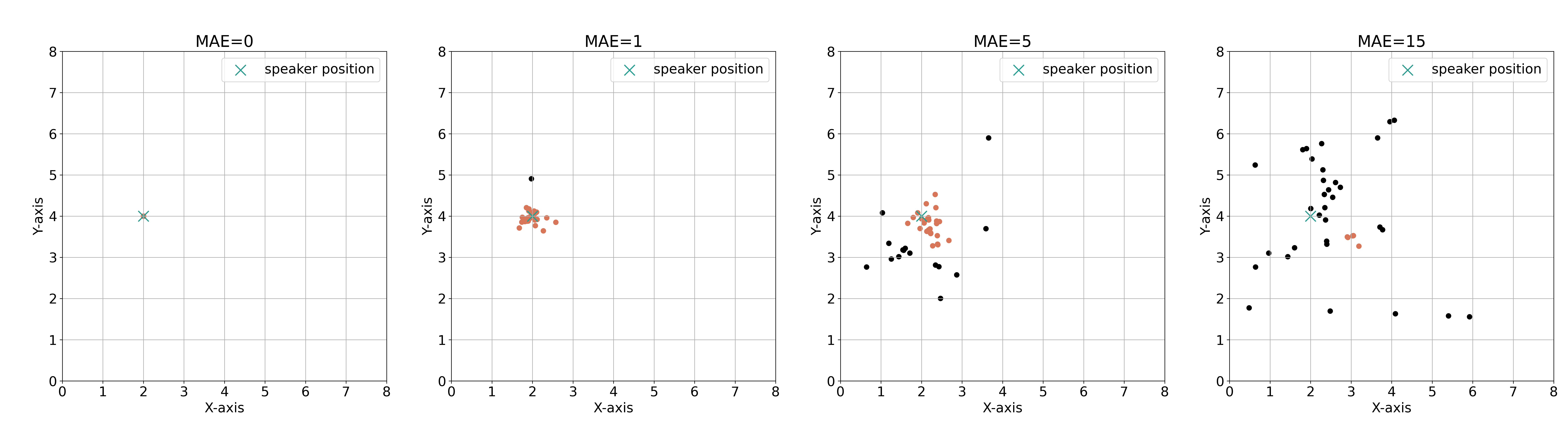}
    \label{subfig:demo_cir1}
  }
  \hfill
  \subfigure[1 speaker, linear arrays]{
    \includegraphics[width=0.92\textwidth]{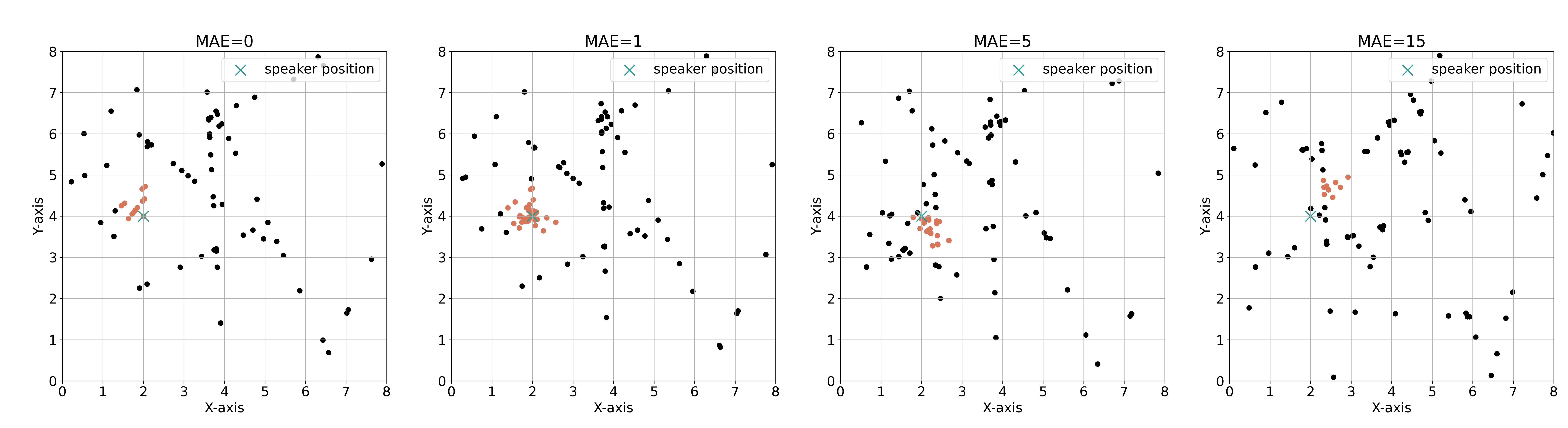}
    \label{subfig:demo_li1}
  }
  \subfigure[2 speakers, circular arrays]{
    \includegraphics[width=0.92\textwidth]{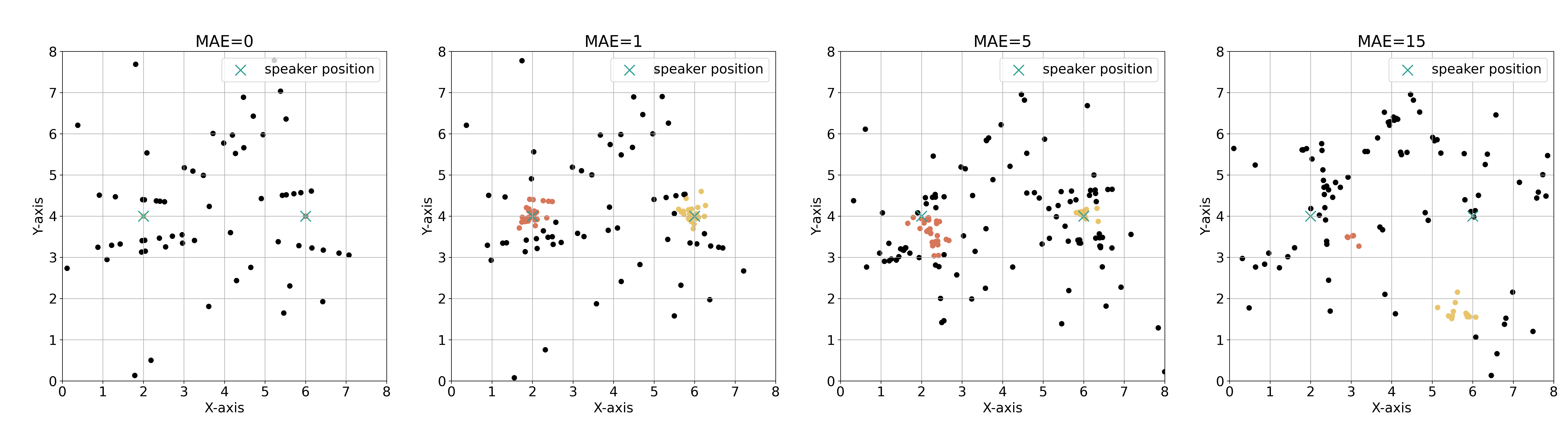}
    \label{subfig:demo_cir2}
  }
  \hfill
  \subfigure[2 speakers, linear arrays]{
    \includegraphics[width=0.92\textwidth]{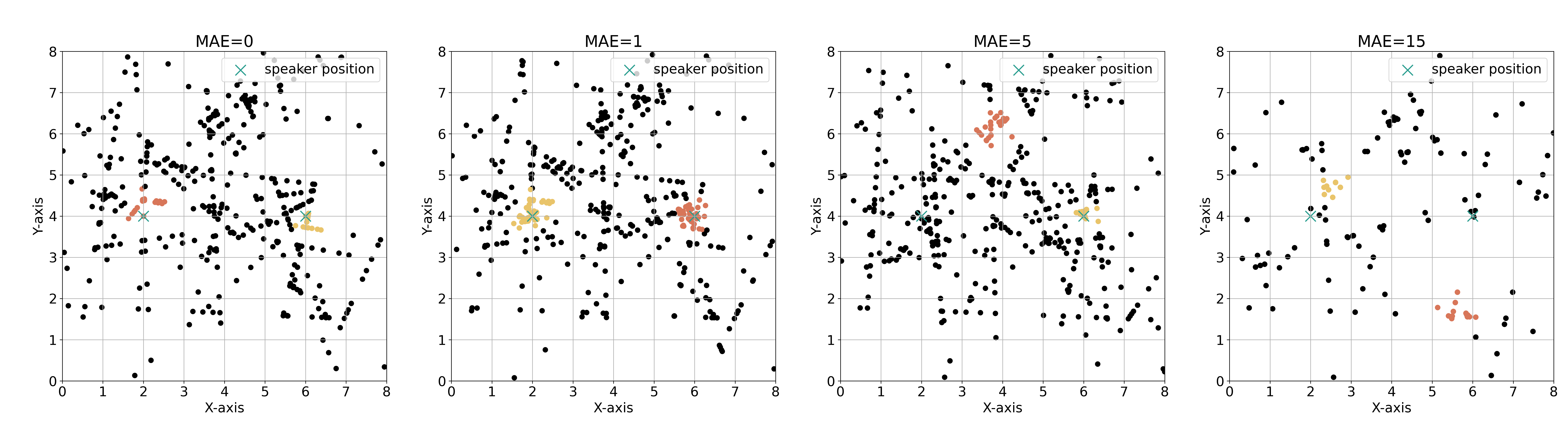}
    \label{subfig:demo_li2}
  }
  \caption{Visualizations of the intersections produced by the pairs of ad-hoc nodes with respect to the MAE levels. The red and yellow points are the selected points by mean-shift clustering for estimating the final 2D speaker locations. }
  \label{fig:demo}
\end{figure*}

Fig. \ref{fig:demo} visualizes the intersections produced by the pairs of ad-hoc nodes with respect to the MAE levels. From Fig. \ref{subfig:demo_cir1}, we see that given a single speaker, the intersections produced by the circular arrays did not contain ghost speakers. When $\mathrm{MAE}=0$, all intersection points converge precisely at the ground-truth position. When the MAE level steadily increases, the intersections gradually disperse. When $\mathrm{MAE}=15$, the MDE produced by the proposed 2D sound source localization system exceeds 1 meter.

From Fig.~\ref{subfig:demo_li1}, we see that numerous ghost speakers emerge when the ad-hoc nodes are linear arrays. When $\mathrm{MAE}=0$, although the correct intersections still converge at the ground-truth position, a few ghost speakers around the clustering center are also taking into the mean-shift clustering, resulting in a slight deviation. As long as the MAE level is not excessively large, the MDE of the 2D localization remains acceptable, which indicates the robustness of the mean-shift clustering against ghost speakers.

From Fig.~\ref{subfig:demo_cir2}, we see that, for the two-source scenario with circular arrays, many ghost speakers that are far from the ground-truth positions are generated. When $\mathrm{MAE}=0$, the 2D localization remains accurate due to the robustness of the mean-shift clustering. However, when the MAE level increases, the MDE of the 2D localization becomes large. When $\mathrm{MAE}=15$, the MDE of the 2D localization for the speaker at $(6, 4)$ exceeds 2 meters.

From Fig.~\ref{subfig:demo_li2}, we see that, for the two-source scenario with linear arrays, the number of ghost speakers is boosted to an unacceptable level, which causes the mean-shift clustering failed to produce reliable 2D localization estimates. For example, when the MAE level is 5 degrees, the estimated position of one speaker is clearly incorrect.

\subsubsection{Quantitative analysis}

\begin{table}[t]
  \centering
  \caption{MDE (in meters) of the 2D sound source localization system with respect to the MAE level at the ad-hoc nodes.}
  \scalebox{0.9}{\begin{tabular}{ccccc}
    \toprule
    \multirow{2}[4]{*}{MAE} & \multicolumn{2}{c}{Circular array} & \multicolumn{2}{c}{Linear array} \\
\cmidrule{2-5}          & 1 speaker & 2 speakers & 1 speaker & 2 speakers \\
    \midrule
    0     & 0.000 & 0.013 & 0.018 & 0.265 \\
    1     & 0.058 & 0.077 & 0.092 & 0.587 \\
    5     & 0.405 & 1.019 & 0.892 & 1.764 \\
    15    & 1.424 & 1.906 & 2.196 & 2.307 \\
    \bottomrule
    \end{tabular}}
  \label{tab:mde_mae}%
\end{table}

We used the simulated single-source and two-source data described in Section~\ref{subsec:data_simu}.
Instead of using the comparison DOA estimation algorithms for the analysis, we just added random perturbation to the ground-truth DOA of each node so as to control the expectation of the MAE levels at the node to be $[0 , 1, 5, 15]$ degrees respectively.

Table~\ref{tab:mde_mae} lists the MDE performance of the 2D speaker localization system with respect to the MAE level at the ad-hoc nodes. From the table, we observe that, when $\mathrm{MAE}=0$, the results are satisfactory. When $\mathrm{MAE}=1$, localizing two speakers with linear arrays becomes challenging due to the interference of a vast number of ghost speakers. When MAE increases to 5 degrees, the proposed system does not work well in all scenarios except the single-source localization scenario using circular arrays. When MAE reaches 15 degrees, the performance becomes fully unacceptable due to the unreliable DOA estimation.

%

\section{Conclusion}\label{sec:conclusion}
In this paper, we have developed a stage-wise deep-learning-based speaker localization method using large-scale ad-hoc microphone arrays. It first applies the CNN-based DOA estimation algorithm at each ad-hoc node. Then, it uses the predicted distributions from the CNN to select reliable nodes. The triangulation method is used to generate numerous rough speaker position estimates from pairs of selected reliable nodes. Finally, mean-shift clustering is applied to the coarse estimates, and the clustering centers are regarded as the accurate 2D position estimates. Extensive comparisons on both simulated and real-world data demonstrate the effectiveness and generalization capabilities of the proposed approach. We also discussed the negative effects of the DOA estimation errors and number of ghost speakers on performance both theoretically and empirically, which needs to be further addressed in the future.


\bibliographystyle{IEEEtran}

\bibliography{Reference}

\end{document}